\documentstyle[psfig,draft,aps]{revtex}
\begin{document}
\draft
\title{Strong Isospin-Breaking Effects 
in $K\to 2\pi$ at Next-to-Leading Order in the
Chiral Expansion}
\author{Carl E. Wolfe\thanks{e-mail: wolfe@niobe.iucf.indiana.edu}}
\address{Nuclear Theory Center, 
Indiana University, Bloomington, IN, 47408}
\author{Kim Maltman\thanks{e-mail: maltman@fewbody.phys.yorku.ca}}
\address{Department of Mathematics and Statistics, York University, \\
4700 Keele St., Toronto, Ontario, Canada M3J 1P3 \\ and }
\address{Special Research Centre for the Subatomic Structure of Matter, \\
University of Adelaide, Australia 5005.}
\maketitle
\begin{abstract}
Strong isospin-breaking (IB) contributions to both the octet and
$27$-plet weak $K\rightarrow \pi\pi$ transitions
are evaluated at next-to-leading order
(NLO) in the chiral expansion.
NLO contributions are shown to significantly reduce the leading
order result for the potentially large contribution to the 
$\Delta I=3/2$ amplitude resulting from strong isospin-breaking
modifications to the weak $\Delta I=1/2$ amplitude.  The
ratio of strong IB $27$-plet to strong IB octet contributions
is found to be small for all decay amplitudes.
Combined with 
recent results on the corresponding electromagnetic contributions,
we find that
the ratio of the intrinsic strengths of octet and
$27$-plet effective weak operators can be taken to be that obtained
from experimental data, analyzed ignoring isospin breaking,
to an accuracy better than of order $\sim 10\%$.  
\end{abstract}

\section{Introduction}
It appears likely that 
the large ratio ($\sim 20$) between octet $\Delta I=1/2$ 
and $27$-plet $\Delta I=3/2$
amplitudes in hyperon and non-leptonic $K$ decay (the so-called
$\Delta I=1/2$ Rule) results from a compounding of long-distance and 
short-distance effects, and that the sources of both effects are
now reasonably well understood.  
QCD dressing, in the regime of scales 
$>1$ GeV$^2$, for which perturbative QCD can be sensibly
employed, contribute a factor of $3-4$ to the 
enhancement\cite{pqcd1loop,pqcd2loop}, while,
in the case of non-leptonic $K$ decay, long-distance effects,
including those of final state interactions (FSI)\cite{buras87,isgur90,kmw91}
also contribute significantly.  Attempts to provide a 
sensible matching of short and long distance effects in a single
theoretical framework now appear likely to account for the full
observed enhancement\cite{buras87,bertolini98,bijnens99,hambye99}.
The neglect of isospin breaking (IB), however, represents
a potential problem for this putative
understanding~\cite{cheng}.  Indeed, since
the ratio of magnitudes of the $\Delta I=1/2$ and
$\Delta I=3/2$ amplitudes is $\sim 20$, IB
at the typical few percent scale
could lead to a ``leakage''
of the large weak $\Delta I=1/2$ transition strength into the
$\Delta I=3/2$ channel with a strength $\sim 20\times$ a few $\%$.  
Were the experimental $\Delta I=3/2$ amplitude to include such
a large contribution, an isospin-conserving (IC) calculation
which reproduced the {\it experimental} ratio of amplitudes
could, in fact, be in error by as much as a factor of $\sim 2$.

At leading order in the chiral expansion,
and for conventional field choices, 
strong IB in $K\rightarrow \pi\pi$ has only two sources:
$\pi^0-\eta$ mixing on the external
$\pi^0$ legs, and IB in the squared $K$ masses
(which produces ``kinematic'' contributions as a result of
the momentum dependence of the
weak transition amplitudes).
At this order, the resulting $\Delta I=1/2$ leakage contribution
represents $\sim 15\%$ of the
observed $\Delta I=3/2$ amplitude\cite{cheng,dgh}.  
Next-to-leading order (NLO) IC corrections are known to be important
for the $I=0$ final state (the $\Delta I=1/2$
transition)\cite{kmw91} and hence are unlikely
to be negligible for the IB corrections.  Some phenomenological
estimates~\cite{etap}, in fact, suggest that they are rather large: 
Ref.~\cite{dgh}, for example, estimates that including the effect of mixing
with the $\eta^\prime$ (a pure NLO effect) raises the $\Delta I=1/2$
leakage contribution to the $\Delta I=3/2$ amplitude
to $35\%$ of the total.  
(In contrast, recent evaluations of electromagnetic
(EM) contributions to the $K\rightarrow\pi\pi$ amplitudes~\cite{cdg} 
find them to represent
few to several percent effects in all three channels, i.e.,
strongly suppressed relative to the naive estimate given above.)
There are, however, other strong-IB-induced NLO contributions
not included in the estimate based only on the effect
of $\eta^\prime$ mixing.  Since the method of effective
chiral Lagrangians~\cite{glchpt} 
(Chiral Perturbation Theory, or ChPT) provides
a straightforward method of evaluating the sum of all such
NLO contributions, we will, in this work, determine the strong
IB contributions to the $K\rightarrow \pi\pi$ amplitudes, including the
leakage contribution of the weak $\Delta I=1/2$
transition, to NLO in ChPT.

\section{The Strong Isospin-Breaking Contributions to 
$K\rightarrow\pi\pi$}

IB has two sources in the Standard Model, electromagnetic (EM)
and strong (due to $m_d\not= m_u$).  EM IB has $I=0,1,2$
components, and hence, in combination with the dominant
$\Delta I=1/2$ octet weak transition operator, produces
contributions to $K\rightarrow\pi\pi$ with $\Delta I=1/2, 3/2$ and 
$5/2$.  These contributions have been recently studied in
Refs.~\cite{cdg}.  Strong IB,
to ${\cal O}(m_d-m_u)$, is,
in contrast, pure $I=1$.
The strong modifications of the basic
$\Delta I=1/2, 3/2$ transitions thus again produce
$\Delta I=1/2, 3/2$ and $5/2$ contributions. 
Due to the factor
of $\sim 20$ difference in the octet
$\Delta I=1/2$ and $27$-plet $\Delta I=3/2$ weak
operator strengths, one would expect
strong IB octet contributions to dominate those
associated with the weak $27$-plet.

In the presence of IB (including the $\Delta I=2$ component of EM,
which couples the $I=0$ and $I=2$ $\pi\pi$ channels)
the analogue of the standard isospin decomposition of
the $K^+\rightarrow\pi^+\pi^0$, 
$K_S\rightarrow\pi^+\pi^-,\pi^0\pi^0$ decay amplitudes, $A_{+0}$,
$A_{+-}$ and $A_{00}$, is~\cite{cdg}
\begin{eqnarray}
\label{isodecomp}
A_{00} &=& \sqrt{2\over 3} A_0-{2\over\sqrt{3}} A_2 
= \sqrt{2\over 3}\vert A_0\vert 
{\rm e}^{i(\Phi_0+\gamma_0)}-{2\over\sqrt{3}}\vert A_2 \vert {\rm e}^
{i(\Phi_2+\gamma_2)}, \cr
A_{+-} &=& \sqrt{2\over 3} A_0 +{1\over\sqrt{3}} A_2=
\sqrt{2\over 3}\vert A_0\vert 
{\rm e}^{i(\Phi_0+\gamma_0)}+{1\over\sqrt{3}}\vert A_2\vert {\rm e}^
{i(\Phi_2+\gamma_2)}, \cr
A_{+0} &=& {\sqrt{3}\over 2} A_2^\prime =
{\sqrt{3}\over 2}\vert A_2^\prime\vert {\rm e}^{i(\Phi_2
+\gamma_2^\prime 
)},
\end{eqnarray}
where the $\Phi_I$ are the {\it strong} $\pi\pi$ (rescattering) phases.
In the absence of $\Delta I=2$ FSI,
$\gamma_2^\prime =\gamma_2$, and $\Phi_I+\gamma_I\equiv \phi_I$ 
should be the physical isospin $I$ $\pi\pi$ scattering phase, $\delta_I$.
In general, $\vert A_2^\prime\vert \not= \vert A_2\vert$ 
as a consequence of EM- and strong-IB-induced
$\Delta I=5/2$ contributions.  If one ignores IB, $A_2^\prime
= A_2$ and $\phi_I=\delta_I$.  

The conventional IC analysis of
$K\rightarrow\pi\pi$ involves first determining
$\vert A_2^\prime\vert$ (assumed equal to $\vert A_2\vert$) from
the $K^+\rightarrow\pi^+\pi^0$ decay rate, 
and then extracting $\vert A_0\vert$
using the IC relation
\begin{equation}
2\vert A_{+-}\vert^2 +\vert A_{00}\vert^2 = 2\vert A_0\vert^2
+2\vert A_2\vert^2 \ .
\label{A0extraction}
\end{equation}
If IB is indeed negligible, then the relative phase,
$\phi =\phi_0 -\phi_2$, of the $I=0$/$I=2$ interference
terms in the two $K_S$ decay rates should 
equal $\delta_0-\delta_2$.
Fitting to the experimental decay rates~\cite{pdg98,devdick}
assuming IC,
one finds
\begin{eqnarray}
\label{expfit}
\vert A_0\vert &=& (4.70\pm 0.01)\times 10^{-4}\;{\rm MeV}\cr
\vert A_2\vert &=& (2.11\pm 0.04)\times 10^{-5}\;{\rm MeV}\cr
\phi &=& 0.98 \pm 0.06\;{\rm rad}.
\end{eqnarray}
The large value of the ratio $\vert A_0/A_2\vert = 22.3$ reflects the
well-known $\Delta I=1/2$ Rule, while
the deviation of the nominal value of $\phi$,
$\phi_{exp}\simeq 56^o$, from $\delta_0-\delta_2=(42\pm 4)^o$ 
presumably reflects the presence of neglected IB contributions.

In general, the two $K_S$ decay
rates depend on {\it three} parameters, $\vert A_0\vert$,
$\vert A_2\vert$, and $\phi$.  Since,
in the presence of $\Delta I=5/2$
IB contributions, $\vert A_2\vert$ can no
longer be determined in $K^+\rightarrow\pi^+\pi^0$,
$\phi$ is not, in fact, experimentally measurable.
The (assumed) IC analysis produces a nominal value, $\phi_{exp}$,
related to the actual value, $\phi$, by
\begin{equation}
\cos\left( \phi_{exp}\right) = {\frac {\vert A_2\vert}
{\vert A_2^\prime\vert}}\cos\left(\phi \right)
+{\frac{\left[ 
\vert A_2^\prime\vert^2 - \vert A_2\vert^2\right]}
{2\sqrt{2}\, \vert A_0\vert\, 
\vert A_2^\prime\vert}}\ .
\label{phaserelation}
\end{equation}
In the presence of $\Delta I=5/2$ transitions, the
coefficient of the first term on the RHS is $\not= 1$, and
the second (small) term is non-zero. 
$\phi_{exp}$ can thus 
differ from $\delta_0-\delta_2$ ,
even if $\Delta I=2$ EM FSI effects are negligible.

We now outline the ingredients needed to
compute the strong IB contributions to the
CP-even $K\rightarrow\pi\pi$ amplitudes in ChPT.

The low-energy representation of
the strong interactions, sufficient to determine effects at
NLO, is given
by the 1-loop effective Lagrangian of Ref.~\cite{glchpt}.  Writing
${\cal L}_S = {\cal L}_S^{(2)}+{\cal L}_S^{(4)}$, 
where the superscripts denote chiral order, and setting the
external vector and axial vector fields (not required for our purposes)
to zero, one has
\begin{eqnarray}
{\cal L}_S^{(2)} &=& {F^2\over 4}{\rm Tr}[\partial_{\mu}U
\partial^{\mu}U^{\dagger}] +
{F^2\over 4}{\rm Tr}[\chi U^{\dagger}+U\chi^{\dagger}]\label{Ltwo} \\
{\cal L}_S^{(4)} &=&L_1({\rm Tr}[\partial_{\mu}U
\partial^{\mu}U^{\dagger}])^2 
+L_2{\rm Tr}[\partial_{\mu}U\partial_{\nu}U^{\dagger}]
\, {\rm Tr}[\partial^{\mu}U\partial^{\nu}U^{\dagger}] 
+L_3{\rm Tr}[\partial_{\mu}U\partial^{\mu}U^{\dagger}
\partial_{\nu}U\partial^{\nu}U^{\dagger}] 
\nonumber \\
&&\ \
+L_4{\rm Tr}[\partial_{\mu}U\partial^{\mu}U^{\dagger}] 
\, {\rm Tr}[\chi U^{\dagger}+U\chi^{\dagger}]
+L_5{\rm Tr}[\partial_{\mu}U\partial^{\mu}U^{\dagger}
(\chi U^{\dagger}+U\chi^{\dagger})] +
L_6({\rm Tr}[\chi U^{\dagger}+U\chi^{\dagger}])^2 \nonumber \\
&&\ \ 
+L_7({\rm Tr}[\chi U^{\dagger}-U\chi^{\dagger}])^2 
+L_8{\rm Tr}[\chi U^{\dagger}\chi U^{\dagger} + 
U\chi^{\dagger}U\chi^{\dagger}]
+H_2{\rm Tr}[\chi\chi^{\dagger}]\label{Lfour}\ ,
\end{eqnarray}
where $\chi = 2B_0 M_q$
(with $M_q$ the quark mass matrix), $U = exp(i\lambda\cdot\pi/F)$
and $\{ L_i\}$, $F$ and $B_0$ are
the usual strong low-energy constants
(LEC's), in the notation of Ref.~\cite{glchpt}, for which we
employ the values found in Ref.~\cite{ecker95}.

The low-energy representation of the CP-even part
of the non-leptonic weak interactions is
given by the Lagrangian, ${\cal L}_W$, of Ref.~\cite{kambor,kmw90}
(or the equivalent reduced forms of Refs.~\cite{ekw93,bpp}, which
take into account constraints associated with the Cayley-Hamilton
theorem and the leading order equation of motion).  We  work with the 
version in which the weak mass term present in the most 
general form of the leading (second) order
part of ${\cal L}_W$, ${\cal L}_W^{(2)}$, has been removed
by a field redefinition, and the NLO part of ${\cal L}_W$, ${\cal L}_W^{(4)}$,
correspondingly modified
(see Ref.~\cite{kmw90} for details).
With ${\cal L}_W^{(2)}=
{\cal L}_{W(8)}^{(2)} + {\cal L}_{W(27)}^{(2)}$, where the subscripts
label the flavor octet and $27$-plet components, respectively, one has
\begin{eqnarray}
{\cal L}_{W(8)}^{(2)}&=&c_2{\rm Tr}\left[\lambda_6D_{\mu}U^\dagger
D^{\mu}U\right] 
\label{LWoctlead} \\
{\cal L}_{W(27)}^{(2)} &=& c_3 t\, \left({\rm Tr}[\tilde{Q}L_{\mu}]
{\rm Tr}[\tilde{Q}L^{\mu}]\right)\ ,
\label{LW27lead}
\end{eqnarray}
where $L_\mu =iU^\dagger D_\mu U$,
$c_2$ and $c_3$ are leading order weak LEC's of order
$G_F$, $D_\mu$ is the covariant derivative (which, for our purposes,
reduces to the ordinary partial derivative),
the matrix $\tilde{Q}$ projects out the flavor octet components of any
trace in which it occurs, and the tensor, $t$, combines two octets
into a $27$-plet.  The explicit forms of $\tilde{Q}$ and
$t$, including those required
to separate the $\Delta I=1/2$ and $3/2$ components of the
$27$-plet, may be found in Ref.~\cite{thesis}.{\begin{footnote}{
Our definition of the ${\cal O}(p^2)$ weak octet operator, and
hence our normalization convention for $c_2$, agrees with
that of Refs.~\cite{kmw91,kambor,kmw90}.  The choice $c_2>0$
conforms to that of
Ref.~\cite{kmw91}, but differs by a sign from that used in Ref.~\cite{thesis}.
Our convention for the tensor $t$ is such that data then requires
$c_3>0$, which differs by a sign from the convention of 
Ref.~\cite{kmw90}, and by both a sign and a factor of $2$ from that
of Refs.~\cite{kmw91,kambor}.  With $c_2>0$, $c_3>0$, our tree-level
amplitudes $A_0$ and $A_2$ are negative.  Our invariant amplitudes 
$A_0$, $A_2$ and $A_2^\prime$
differ by a factor of $-\sqrt{3}$ from those of
Ref.~\cite{cdg}.  This can be understood from a comparison of
the expressions for the amplitudes, provided one takes into account
the fact that the neutral
$K$ decay amplitudes of Refs.~\cite{cdg} refer to $K^0$ decays,
whereas ours, following the notation of Ref.~\cite{kmw91}, 
refer to $K_S$ decays.
}\end{footnote}}

For the NLO weak contributions one has~\cite{kmw90},
\begin{equation}
\label{LWnlo}
{\cal L}^{(4)}_W = \sum_{i=1}^{48}E_i{\cal O}^{(8)}_i +
\sum_{i=i}^{34}D_i{\cal O}^{(27)}_i
\end{equation}
where the $E_i$ and $D_i$ are the
weak NLO octet and $27$-plet LEC's
(which have an implicit proportionality to $c_2,c_3$, respectively).  
The corresponding renormalized LEC's are denoted $E_i^r$
and $D_i^r$.  Their relations to the $E_i$ and $D_i$
are given in Ref.~\cite{kmw90} 
{\begin{footnote}{We concur with Ref.~\cite{bpp} in requiring
an overall difference in the sign of the divergent parts of
all $27$-plet LEC's, as compared to the
results of Ref.~\cite{kmw90}.}\end{footnote}}.
Explicit expressions
for the operators ${\cal O}^{(n)}_i$ may be found
in Ref.~\cite{kmw90}.
Use of the Cayley-Hamilton theorem and the equation of motion
allows one to remove certain of the terms in Eq.~(\ref{LWnlo}),
as explained in Section 3 of Ref.~\cite{kmw90}.  We work
with a form in which the former
constraint has been used to eliminate $E_{14}^r$, and the
LEC's $E_{10}^r$ through $E_{13}^r$ modified accordingly.
(The constraint also allows elimination of $E_{44}^r$,
$E_{45}^r$, and $D_{32}^r$, but this is irrelevant
for our purposes since the corresponding operators
do not contribute to $K\to\pi\pi$ at NLO.)
The reader should bear in mind that,
in employing the GNC model\cite{holdom92,cameron} below
to make estimates for the weak LEC's, one must also impose
this constraint, which has not been implemented in 
Ref.~\cite{cameron}.
\begin{footnote}{Ref.~\cite{cameron} also employs a form of the
strong NLO Lagrangian in which, in contrast to the conventional
form given above, operators which could be omitted as a
consequence of the leading order equation of motion have not been removed.
In order to employ the results of Ref.~\cite{cameron} 
one must, therefore, first remove those operators,
and then make the corresponding changes to the weak LEC's.  These
modifications affect only the values of $E_{15}$,
$E_{32}$, $E_{33}$, $D_5$, and $D_{19}$ of Ref.~\cite{cameron}. }
\end{footnote}

An alternate choice of operator basis for the NLO weak octet
Lagrangian is that given in Ref.~\cite{ekw93}.
When we employ the weak deformation/factorization model (FM)
estimates of Ref.~\cite{ekw93} for the weak LEC's,
we will work with the reduced set of octet operators,
$W^{(8)}_i$, and corresponding LEC's, $N_i$, defined in that
reference. For the corresponding weak $27$-plet operators we follow
Ref.~\cite{bpp}, denoting the operators by $\tilde{O}_i^{(27)}$
and the LEC's by $\tilde{D}_i$.  The renormalized LEC's are written,
in obvious notation, $N_i^r$ and $\tilde{D}_i^r$,
respectively.

Certain combinations of the weak LEC's were determined
in Ref.~\cite{kmw91} by neglecting IB and fitting the calculated
${\cal O}(p^4)$ amplitudes for $K\rightarrow\pi\pi$ and 
$K\rightarrow\pi\pi\pi$ 
to experimental data. (Sufficient data exists to allow such an IC
fit, provided one neglects contributions suppressed by a
relative factor of $m_{\pi}^2/m_K^2$\cite{kmw91}.)
Since all IC octet (respectively, $27$-plet) contributions are proportional to 
$c_2$ (respectively, $c_3$), 
the presence of IB contributions can be accommodated
in the fit by rescalings of $c_2$ and $c_3$.  One, of course,
expects a small rescaling for $c_2$, but potentially significant
rescaling for $c_3$.
As can be seen from the results below,
the LEC combinations entering the IB contributions to 
$A_0$, $A_2$ and $A_2^\prime$ are such that
the total number of linearly independent IC and IB LEC combinations
exceeds the existing number of $K\rightarrow\pi\pi$ and 
$K\rightarrow\pi\pi\pi$ observables, making
an experimental determination of the new IB LEC values
impossible.  It is, therefore, necessary to estimate
their values using models.  
We employ two models for this purpose, each 
representing the extension of a model successful in reproducing
the empirical values of the strong LEC's.

In the first of these models, the FM~\cite{ekw93},
a rescaled version of the factorization of
the four-quark currents into products of two-quark currents is employed, the
LEC contributions to the latter being given by
resonance saturation
(see Ref.~\cite{egpdr89} for the modelling of
the strong LEC's, and Refs.~\cite{ekw93,bpp} for an explicit discussion
of the relation to the weak LEC's).
In the second model, the gauged non-local
constituent quark (GNC) model (a chiral quark model with a
self-energy, $\Sigma_A (q^2)$, modelled
using a parameter, $A$, describing the rapidity of
onset of the constituent mass with $q^2$),
the effective Lagrangians for the pseudoscalars are 
generated by integrating out the quark fields~\cite{holdom92,cameron}.
Values of $A$ between $1$ and $3$ give good fits to the strong
LEC's. We will employ the FM and the GNC (with both $A=1$ and $A=3$),
using the difference in the estimates so obtained 
to provide a measure of
the uncertainty associated with
the model dependence of the weak LEC values. 

Using the expressions above for the weak and strong effective
Lagrangians, it is straightforward to compute the desired strong
IB contributions to $A_0$, $A_2$ and $A_2^\prime$.
The leading (${\cal O}(p^2)$) contributions are given in Table \ref{table1}.
\begin{footnote}{The IB $\pi$-$\eta$ mixing and kinematic contributions
turn out to exactly cancel for $K\rightarrow\pi^0\pi^0$, in both
the octet and $27$-plet cases.  The mixing contribution is, of course,
absent for the $K\rightarrow\pi^+\pi^-$ amplitude.}
\end{footnote} 
The NLO contributions are obtained by
evaluating the graphs of Figs.~1(b)-(h).  
The notation for the Figures is as follows: internal lines
represent any of the members of the pseudoscalar octet,
solid circles the ${\cal O}(p^2)$ strong vertices,
open circles the ${\cal O}(p^4)$ strong vertices, solid squares the
${\cal O}(p^2)$ weak vertices, and the open square of
Fig.~1(h), any of the ${\cal O}(p^4)$ weak vertices.  
Expressions relating the isospin-pure, non-diagonally-propagating 
$\pi^3$ and $\pi^8$ fields to the renormalized,
mixed-isospin, diagonally-propagating $\pi^0$ and $\eta$ fields at NLO,
which are required to handle the effects of $\pi$-$\eta$ mixing
at this order, are taken
from Ref.~\cite{kmmix}, while expressions for the ${\cal O}(m_d-m_u)$
contributions to the wavefunction renormalization factors may be found
in Ref.~\cite{thesis}.

Since the expressions for the IB
parts of the one-loop graphs (Figures 1(b)-(g)) 
are rather lengthy and unilluminating, we record here
only their numerical values.
\begin{footnote}{Expressions for the octet
one-loop IB contributions may be 
found in Appendix B of Ref.~\cite{thesis}; those for the $27$-plet
will be reported elsewhere~\cite{promisespromises}}.\end{footnote}
The results (including the 
strong LEC contributions to the NLO mixing and
wavefunction renormalizations (Fig. 1(c)),
of which the $L_7$ term, which reflects the $\eta^\prime$
mixing contribution at this order, is a part) 
are given in
Table \ref{table2}.  These results correspond,
for definiteness, to the scale $\mu^2=m_\eta^2$, and 
are given in the form $\delta^{(f)}A_i/c^{(f)}$,
where $(f)=(8)$ or $(27)$ labels the flavor of the weak transition
operator, $c^{(8)}\equiv c_2$, $c^{(27)}\equiv c_3$, and
$A_i=A_0$, $A_2$ or $A_2^\prime$ (these combinations are
independent of the specific values of the weak LEC's $c_2$ and $c_3$).
The scale dependence of each such sum must, of course, cancel 
that of the corresponding weak LEC combination.
Collectively, the finiteness and scale independence of 
each of the three $K\rightarrow\pi\pi$ decay amplitudes
provides a powerful cross-check on the calculations.

The weak LEC (counterterm) contributions,
corresponding to Fig.~ 1(h), are given by
\begin{eqnarray}
\label{weakct}
\left[ \delta^{(8)} A_0\right]_{LEC} &=& - \left({\frac{\sqrt{6}B_0(m_d-m_u)}
{9F^3}}\right)\left(m_K^2J_1-m_{\pi}^2J_2 \right) 
\nonumber \\
\left[ \delta^{(8)} A_2\right]_{LEC} &=& 
\left[ \delta^{(8)} A_2^\prime \right]_{LEC} =
\left({\frac{2B_0(m_d-m_u)}{\sqrt{3}F^3}}\right)\left( m_K^2J_3-m_{\pi}^2J_4
\right) \nonumber \\
\left[ \delta^{(27)}A_0\right]_{LEC} &=& -\left({\frac{\sqrt{2}B_0(m_d-m_u)}
{4\sqrt{3}F^3}}\right)\left( m_K^2 K_1+m_{\pi}^2 K_2\right) \nonumber \\
\left[\delta^{(27)}A_2\right]_{LEC} &=& \left(
{\frac{B_0(m_d-m_u)}{4\sqrt{3}F^3}}
\right)\left( m_K^2 K_3+m_{\pi}^2 K_4\right) \nonumber \\
\left[\delta^{(27)}A_2^{\prime}\right]_{LEC} &=& \left({\frac{B_0(m_d-m_u)}
{2\sqrt{3}F^3}}\right)\left( m_K^2 K_5+m_{\pi}^2 K_6\right) 
\end{eqnarray}
where, in the basis of Ref.~\cite{kmw90},
\begin{eqnarray}
\label{iblec}
J_1 &=&-12E_1^r+24E_3^r+36E_4^r-12E_5^r+21E_{10}^r+9E_{11}^r+36E_{12}^r
+15E_{15}^r-72E_{32}^r-48E_{33}^r-24E_{34}^r
\nonumber \\
&& \quad +30E_{37}^r +30E_{38}^r \nonumber \\
J_2 &=& -60E_1^r-36E_2^r+12E_3^r+36E_4^r+48E_5^r+33E_{10}^r-12E_{11}^r
+36E_{12}^r
+18E_{13}^r+9E_{15}^r-72E_{32}^r
\nonumber \\
&& \quad +96E_{34}^r +24E_{35}^r+24E_{36}^r+18E_{37}^r+18E_{38}^r-48E_{39}^r
-48E_{40}^r \nonumber \\
J_3 &=&-4E_1^r+8E_3^r+12E_{4}^r-4E_5^r+E_{10}^r+3E_{11}^r+12E_{12}^r
-E_{15}^r-24E_{32}^r-16E_{33}^r-8E_{34}^r-2E_{37}^r-2E_{38}^r \nonumber \\
J_4 &=& -2E_1^r+10E_3^r+12E_4^r-8E_5^r+2E_{10}^r+5E_{11}^r+12E_{12}^r
-24E_{32}^r-24E_{33}^r-16E_{34}^r-4E_{35}^r-4E_{36}^r \nonumber \\
&&\quad +8E_{39}^r+8E_{40}^r \nonumber \\
K_1 &=& 208D_1^r+10D_4^r-10D_5^r-66D_6^r+32D_7^r+20D_{22}^r+20D_{23}^r 
\nonumber \\
K_2 &=& -144D_1^r+32D_2^r+30D_4^r+2D_5^r+50D_6^r+16D_7^r-64D_{19}^r
-32D_{20}^r -32D_{21}^r-4D_{22}^r-4D_{23}^r \nonumber \\
&&\quad +128D_{24}^r+128D_{25}^r \nonumber \\
K_3 &=& -64D_1^r-28D_4^r+28D_5^r+12D_6^r-32D_7^r-56D_{22}^r-56D_{23}^r 
\nonumber \\
K_4 &=& -48D_1^r-32D_2^r-12D_4^r-20D_5^r+16D_6^r-16D_7^r
+64D_{19}^r+32D_{20}^r+32D_{21}^r+40D_{22}^r+40D_{23}^r \nonumber \\
&&\quad -128D_{24}^r-128D_{25}^r \nonumber \\
K_5 &=& -32D_1^r+11D_4^r-11D_5^r+16D_6^r+4D_7^r+22D_{22}^r+22D_{23}^r 
\nonumber \\
K_6 &=& 16D_1^r+4D_2^r-6D_4^r+10D_5^r-12D_6^r+2D_7^r
-8D_{19}^r-4D_{20}^r-4D_{21}^r-20D_{22}^r-20D_{23}^r+16D_{24}^r+16D_{25}^r
\end{eqnarray}
while in that of Ref.~\cite{ekw93,bpp},
\begin{eqnarray}
\label{iblececker}
J_1 &=& {c_2}\left[ 7N_5^r+6N_6^r+4N_8^r+5N_9^r-4N_{10}^r-8N_{12}^r-12N_{13}^r
\right ] /F^2 \nonumber \\
J_2 &=& {c_2}\left[ 11N_5^r+6N_6^r+6N_7^r-2N_8^r+3N_9^r-20N_{10}^r-12N_{11}^r
-4N_{12}^r-12N_{13}^r \right ] /F^2 \nonumber \\
J_3 &=& {c_2}\left[ N^r_5+6N^r_6
-2N^r_8-N^r_9-4N^r_{10}-8N^r_{12}-12N^r_{13}\right] /F^2 \nonumber\\
J_4 &=& {c_2}\left[ 2N^r_5+6N^r_6+N^r_8
-2N^r_{10}-10N^r_{12}-12N^r_{13}\right] /F^2 \nonumber \\
K_1 &=& {-2c_3}\left[-104\tilde{D}_1^r-5\tilde{D}_4^r
+5\tilde{D}_5^r+33\tilde{D}_6^r-16\tilde{D}_7^r\right] /F^2 \nonumber \\
K_2 &=& {-2c_3}\left[ 72\tilde{D}_1^r+16\tilde{D}_2^r
-15\tilde{D}_4^r-\tilde{D}_5^r-25\tilde{D}_6^r-8\tilde{D}_7^r\right] 
/F^2 \nonumber \\
K_3 &=& {4c_3}\left[ -16\tilde{D}_1^r-7\tilde{D}_4^r
+7\tilde{D}_5^r+3\tilde{D}_6^r-8\tilde{D}_7^r\right] /F^2 \nonumber \\
K_4 &=& {4c_3}\left[ -12\tilde{D}_1^r+8\tilde{D}_2^r
-3\tilde{D}_4^r-5\tilde{D}_5^r+4\tilde{D}_6^r-4\tilde{D}_7^r\right] 
/F^2 \nonumber \\
K_5 &=& {c_3}\left[ 
-32\tilde{D}_1^r+11\tilde{D}_4^r
-11\tilde{D}_5^r+16\tilde{D}_6^r+4\tilde{D}_7^r\right] /F^2 \nonumber \\
K_6 &=& {c_3}\left[ 16\tilde{D}_1^r-4\tilde{D}_2^r
-6\tilde{D}_4^r+10\tilde{D}_5^r-12\tilde{D}_6^r+2\tilde{D}_7^r\right]
/F^2\, .
\end{eqnarray}
It is worth commenting that, although the $J_4$ contribution
to $\left[ \delta^{(8)} A_2\right]_{LEC}$
is suppressed by a factor of $m_\pi^2/m_K^2$ relative
to that involving $J_3$, the ratio of the two contributions in
fact ranges between $0.3$ and $0.6$ for the models discussed.
One should, therefore, reserve some caution for 
the procedure of neglecting LEC contributions 
to the $K\rightarrow\pi\pi$ and $K\rightarrow\pi\pi\pi$
amplitudes which are suppressed by
$m_\pi^2/m_K^2$. 

\section{Numerical Results and Conclusions}
Our numerical results are based on the following input:
$\pi$ and $K$ masses and decay constants from Ref.~\cite{pdg98};
strong NLO LEC's from Ref.~\cite{ecker95}; weak LEC values
from the models noted above; and
\begin{equation}
B_0(m_d-m_u) = \left({\frac{m_d-m_u}
{m_d+m_u}}\right) m_{\pi}^2 =
5248\pm 674\;{\rm MeV}^2,
\label{ibvalues}
\end{equation}
which follows from Leutwyler's determination of the light quark
mass ratios~\cite{leut96}.  

In determining the rescaling of
the weak LEC's, $c_2$ and $c_3$, from their IC values,
we include not only our strong octet and
$27$-plet IB contributions, but also
the EM IB contributions, as determined in the most 
constraining (dispersive) version of the analysis
of Refs.~\cite{cdg}.  The difference in the magnitudes of these rescalings
determines the error in the extracted value of the
ratio, $c_3/c_2$, of weak $27$-plet to weak octet operator strengths
made by neglecting IB effects in the analysis of experimental data.
The fitted values of $c_2$ and $c_3$,
together with the ratio 
\begin{equation}
R_{IB}\equiv {\frac{c_3/c_2}{c_3^{IC}/c_2^{IC}}}\ ,
\label{IBeffect}
\end{equation}
which quantifies this error,
are given in Table \ref{table3},
where the IC fit values of $c_2$ and $c_3$ 
have also been included for comparison.
Note that a value of
$R_{IB}<1$ implies that the 
ratio of $\Delta I=1/2$ to $\Delta I=3/2$ operator
strengths is {\it larger} than would be obtained in an IC analysis.
After including the quoted errors on the EM contributions from
Ref.~\cite{cdg}, we find
\begin{equation}
R_{IB}=0.963\pm 0.029\pm 0.010 \pm 0.034\ ,
\label{c2c3effect}
\end{equation}
where the first error reflects the model dependence associated
with the ${\cal O}(p^4)$ weak LEC values, the
second the uncertainty in $B_0(m_d-m_u)$, and the third the uncertainty
in the EM contributions.  The ratio $c_2/c_3$ can thus be taken
to be that obtained in an IC analysis to an accuracy of better
than $\sim 10\%$.

To understand the reason for this rather small IB
shift, it is useful to examine separately the octet, $27$-plet
and EM IB contributions to the $K\rightarrow\pi\pi$
amplitudes.  We denote by $\delta^{(s)}A_k$
the IB contribution to 
$A_k$, where $A_k$ is any of
$A_0$, $A_2$, and $A_2^\prime$, and $(s)=(8)$, $(27)$ or
$(EM)$ labels the source of IB.
The results for the $\delta^{(s)}A_k$, 
are given in Table \ref{table4}.  The EM results and
errors are those of
Refs.~\cite{cdg}, adapted to our conventions.
The errors on the real parts of the strong IB 
contributions correspond to the range of values of
the weak LEC contributions obtained
from the different models above combined in quadrature
with the error associated with the uncertainty in
$B_0(m_d-m_u)$; the former turns out to be the
dominant source of error.

A number of features of the results are worth further comment.
First, in all cases the IB $27$-plet contributions are
a factor of $\sim 20$ smaller than than the IB octet,
compatible with naive estimates based on the relative
size of the $\Delta I=1/2$ and $\Delta I=3/2$ weak
operator strengths.  
Second, the EM contributions are 
of order $\sim 50\%$ of the octet IB contributions for 
$A_0$ and $A_2^\prime$, and of order $\sim 80\%$ for $A_2$,
the two contributions adding constructively for 
$A_0$ and $A_2$, but destructively for $A_2^\prime$.
Third, while in all cases the strong IB contributions add constructively to
the IC contributions, the EM contributions add constructively
for $A_0$ and $A_2$, but destructively for $A_2^\prime$.
These features ensure that $\vert A_2\vert /\vert A_2^\prime\vert >1$,
an effect which tends to make the nominal phase, $\phi_{exp}$,
{\it smaller} than the actual phase difference, $\phi$.  
Because the IB $27$-plet contributions are, as expected, 
small, this effect (associated with the presence of a $\Delta I=5/2$
contribution in the $K\rightarrow\pi\pi$ amplitudes) is almost totally
dominated by the EM component.  In fact, as one can see from the
near equality of $\delta^{(27)}A_2$ and $\delta^{(27)}A_2^\prime$, 
the $27$-plet-induced $\Delta I=5/2$ 
component is strongly
suppressed, in contrast to the situation for the EM contributions.
As a result, though the $27$-plet IB contribution to each of
$A_2$ and $A_2^\prime$ is at the $\sim 10\%$ level of the
corresponding EM contribution, it has been reduced to the $1/2\%$ level
when one considers $\vert A_2\vert -\vert A_2^\prime\vert$.

Let us return to the question of the IB modification  
of $c_3/c_2$, the ratio which parametrizes the
$\Delta I=1/2$ rule enhancement in the low energy effective theory.
We have seen above that the IB effect is, in fact, quite modest.
It is now possible to see why it is that this is the case.
The results of Table \ref{table3} show that, as expected,
$c_2$ is only slightly
modified (at the $\sim 1\%$ level) by IB effects.  
The ratio $c_3/c_3^{IC}$ is, however, much closer to $1$ than
the $15\%$ deviation produced by including only the leading
order strong IB octet contributions.  This decrease in the IB
effect on $c_3$ has two sources.  First, as can be 
seen from Table \ref{table4},
there is a significant cancellation between the octet and
EM IB contributions to $A_2^\prime$, which quantity dominates
the determination of $c_3$.  Second, this cancellation is
facilitated by the fact that
the ${\cal O}(p^2)$ and
${\cal O}(p^4)$ octet leakage contributions add destructively.
This latter feature might seem unnatural given the observation
that $\eta^\prime$ mixing is expected to {\it increase}
the leading order octet IB effect, but there is, in fact, 
a natural reason why this is not the case.
In the strong interaction part of the 
low energy effective theory, the effects of the $\eta^\prime$
are encoded entirely in the LEC $L_7^r$.  A contribution proportional
to $L_7^r$, associated with the effects of mixing on the external $\pi^0$
legs, is, of course, present in the results above, and
indeed, on its own, would serve 
to significantly increase the leading order result. 
However, as can be seen from Eqs.~(15)-(17) of Ref.~\cite{kmmix},
the LEC contributions to the relevant mixing angles occur
in the combination $3L_7^r+L_8^r$, for which, empirically, 
there is an almost complete
cancellation between the $L_7^r$ and $L_8^r$ terms.
\begin{footnote}{This observation has also been made
in the context of an
estimate of NLO mixing contributions to the IB correction, $\Omega_{IB}$, 
in $\epsilon^\prime /\epsilon$\cite{emnp99}.  A useful discussion
of the resonance interpretation of the $L_8^r$ contribution
can be found in that reference.}\end{footnote}
The cumulative effect of {\it all} NLO corrections, including the strong LEC
corrections just discussed, is, in fact, to lower the magnitude
of the leading order results; the estimate based only on the
inclusion of $\eta^\prime$ mixing effects thus turns out to
be misleading.  One of the great advantages of the ChPT approach is that it
allows one, in a straightforward manner, to include all contributions
of a given chiral order which occur in the Standard Model.

We conclude with a brief comment about the relation of the nominal
phase, $\phi_{exp}$, and the actual relative phase, $\phi$, between the
$I=0$ and $I=2$ components of the two $K_S$ amplitudes.  In order to
fully explicate the phase question, one would require both a determination
of the IB contributions to the (in the presence of EM, coupled channel)
$\pi\pi$ scattering phases, and a determination, and subtraction,
of non-$\pi\pi$-scattering IB effects in the processes in which the
$\pi\pi$ phases are nominally measured.  Such expressions are not
currently available, and a determination of them 
is beyond the scope of this paper.  Without such expressions, however,
the relation between $\phi$ and the nominally determined experimental
$I=0$ and $I=2$ $\pi\pi$ phases is subject to IB corrections
whose size is not, at present, known.
In addition, one should bear in mind that the experimental
data has yet to have applied to it the detector-dependent
IR correction factor present in the
expression for the $K_S\rightarrow\pi^+\pi^-$ 
cross-section (see Ref.~\cite{cdg} for a discussion of this point).  
Since the difference
$\vert A_{+-}\vert^2 -\vert A_{00}\vert^2$, from which
the interference term which determines $\phi_{exp}$ is obtained,
is $\sim 10\%$ of the individual terms,
even a $1\%$ IR correction can have a sizeable numerical impact.
While the problems just discussed mean that uncertainties exist, both
in our knowledge of the
relation of $\phi$ to the measured $\pi\pi$ scattering phase
difference, and in the experimental determination of $\phi_{exp}$,
our results, combined with those of Ref.~\cite{cdg}, allow us
to quantify the deviation of $\phi_{exp}$ from $\phi$
resulting from the presence of $\Delta I=5/2$ strong and EM IB
effects.  We find, for the coefficient of $\cos (\phi )$ in
Eq.~(\ref{phaserelation}),
\begin{equation}
{\frac{\vert A_2\vert}{\vert A_2^\prime \vert}}=1.094\pm 0.039\ .
\label{ratioa2}
\end{equation}
The second term in Eq.~(\ref{phaserelation}) is then
$-0.0015\pm 0.0006$; its effect is thus tiny, and in any case,
swamped by the error on $\vert A_2\vert /\vert A_2^\prime \vert$.
As an example of the magnitude of the resulting effect, note
that, were $\phi$ to be $42^o$, one would then obtain
$\phi_{exp}=35.8\pm 2.9^o$.  Recall that this effect is almost
entirely EM in origin.  The sign of the EM contributions is thus
such as to significantly exacerbate the existing phase 
discrepancy.
\acknowledgements
CEW would like to thank J. Kambor for making a copy of his Ph.D. thesis 
available and acknowledges both the support and hospitality
of the Department of Physics and Astronomy at York University,
during the course of his thesis work there, and the partial support of 
the United States Department of Energy (contract DE-FG0287ER-40365).  
KM acknowledges the ongoing support of the Natural
Sciences and Engineering Research Council of Canada, and the
hospitality and support of the Special Research Centre for the Subatomic
Structure of Matter at the University of Adelaide.

\vfill\eject

\begin{table}
\caption{The ${\cal O}(p^2)$ contributions to $A_0$, $A_2$ and
$A_2^\prime$, in units of $B_0(m_d-m_u)/F^3$.
The label $(f)$ denotes the flavor of the basic weak transition
($(8)$ for octet, $(27)$ for $27$-plet).}\label{table1}
\begin{tabular}{cccc}
$(f)$&$\delta^{(f)}A_0$&$\delta^{(f)}A_2$&$\delta^{(f)}A_2^\prime$ \\
\tableline
$(8)$&$-\sqrt{\frac{2}{3}}\, c_2$&$-\sqrt{\frac{1}{3}}\, c_2$&
$-\sqrt{\frac{1}{3}}\, c_2$ \\
$(27)$&$-2\sqrt{\frac{2}{3}}\, c_3$&$-2\sqrt{\frac{1}{3}}\, c_3$&
${\frac{1}{2}}\sqrt{\frac{1}{3}}\, c_3$ \\
\end{tabular}
\end{table}

\begin{table}
\caption{Octet and $27$-plet contributions to
$A_0$, $A_2$ and $A_2^\prime$ corresponding to the graphs of
Figures 1(b)-(g).  The quantities
$c^{(8)}\equiv c_2$ (for the octet case) and 
$c^{(27)}\equiv c_3$ (for the $27$-plet case) have been
factored out, for the reasons described in the text.
The entries correspond to the renormalization scale
$\mu^2=m_\eta^2$, and are in units of MeV$^{-1}$.  
As in Table 1, $(f)$ represents the flavor
of the weak transition operator.
The fitted $c_2$ and $c_3$ values, needed in order to determine the actual
numerical values of the corresponding contributions to the
$K\rightarrow\pi\pi$ amplitudes, are given in 
Table \ref{table3}.}\label{table2}
\begin{tabular}{cccc}
$(f)$&$\delta^{(f)}A_0/c^{(f)}$
&$\delta^{(f)}A_2/c^{(f)}$&$\delta^{(f)}A_2^\prime /c^{(f)}$ \\
\tableline
$(8)$&$0.00185-0.00538\, {\rm i}$&$0.00091+0.00078{\rm i}\, $&
$0.00091+0.00078\, {\rm i}$ \\
$(27)$&$-0.00803-0.0119\, {\rm i}$&$0.0181+0.0110\, {\rm i}$&
$-0.0239-0.00775\, {\rm i}$ \\
\end{tabular}
\end{table}

\begin{table}
\caption{Fitted values of $c_2$ and $c_3$ in units of MeV$^2$, and
the corresponding values of $R_{IB}$.  IC labels the IC fit,
while the three IB cases correspond to the three models for the
weak LEC's described in the text.  The results quoted here correspond
to central values of both $B_0(m_d-m_u)$, as given
in Eq.~(\ref{ibvalues}), and the EM contributions, as given in
Ref.~[12]. }\label{table3}
\begin{tabular}{cccc}
Fit type&$c_2$&$c_3$&$R_{IB}$ \\
\tableline
IC&$5.43\times 10^{-4}$&$7.23\times 10^{-6}$&$1.000$ \\
IB, GNC $A=1$&$5.38\times 10^{-4}$&$6.91\times 10^{-6}$&$0.965$ \\
IB, GNC $A=3$&$5.37\times 10^{-4}$&$7.09\times 10^{-6}$&$0.992$ \\
IB, FM&$5.40\times 10^{-4}$&$6.71\times 10^{-6}$&$0.934$ \\
\end{tabular}
\end{table}

\begin{table}
\caption{The strong IB octet, strong IB $27$-plet, and EM IB
contributions to $A_0$, $A_2$ and $A_2^\prime$.  The notation
is as described in the text.
Entries are in units of
$10^{-6}\ {\rm MeV}$.  To understand the scale of
the effects, recall that the IC fit
yields $\vert A_0\vert =4.7\times 10^{-4}\ {\rm MeV}$
and $\vert A_2\vert = \vert A_2^\prime\vert =
2.1\times 10^{-5}\ {\rm MeV}$, and note that
the ${\cal O}(p^2)$ octet IB contributions to
$A_0$ and $A_2 = A_2^\prime$ are $-3.4\times 10^{-6}$ MeV
and $-2.4\times 10^{-6}$ MeV, respectively. }\label{table4}
\begin{tabular}{llll}
Source&$\delta^{(s)}A_0$&$\delta^{(s)}A_2$&
$\delta^{(s)}A_2^\prime$ \\ 
\tableline
$(8)$&$(-4.11\pm 1.22)-(2.89\pm 0.37){\rm i}$&$(-1.56\pm 0.63)
+(0.42\pm 0.05){\rm i}$&
$(-1.56\pm 0.63)+(0.42\pm 0.05){\rm i}$ \\
$(27)$&$(-0.28\pm 0.07)-(0.08\pm 0.01){\rm i}$&$(-0.08\pm 0.05)
+(0.07\pm 0.01){\rm i}$&
$(-0.07\pm 0.02)-(0.05\pm 0.01){\rm i}$ \\
$(EM)$&$(-2.17\pm 0.50)+(0.61\pm 0.02){\rm i}$&$(-1.27\pm 0.40)
-(1.28\pm 0.02){\rm i}$&$(0.70\pm 0.73)-(0.07\pm 0.04){\rm i}$ \\
\tableline
Total&$(-6.56\pm 1.32)-(2.36\pm 0.37){\rm i}$&
$(-2.91\pm 0.75)-(0.79\pm 0.05){\rm i}$&
$(-0.93\pm 0.96)+(0.30\pm 0.06){\rm i}$
\end{tabular}
\end{table}

\vfill\eject

\begin{figure}
\centering{\
\psfig{figure=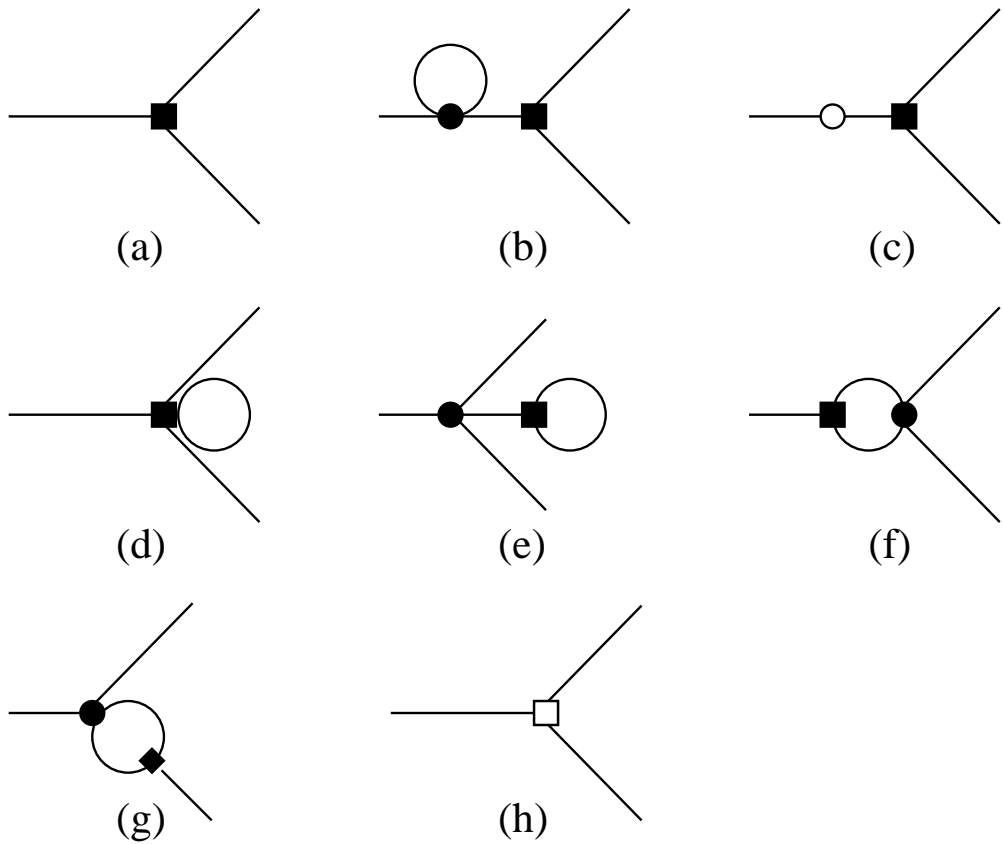}}
\vskip 0.5in
\caption{Feynman diagrams for $K\to\pi\pi$ up to ${\cal O}(p^4)$ in 
the chiral expansion.  Closed circles represent ${\cal O}(p^2)$ strong 
vertices,
open circles ${\cal O}(p^4)$ strong vertices, closed boxes ${\cal O}(p^2)$ weak
vertices, and open boxes ${\cal O}(p^4)$ weak vertices.  No one-line weak 
tadpoles occur because, in the weak effective Lagrangian employed, they have
already been rotated away. Figures (b) and (c) should be understood
to represent collectively the strong dressing on all the 
external lines.}
\label{diagrams}
\end{figure}

\vfill\eject

\end{document}